# Collective Near-Field Thermal Emission from Polaritonic Nanoparticle Arrays

Eric Tervo,[1] Zhuomin Zhang,[1] and Baratunde Cola[1,2,*]

[1] *George W. Woodruff School of Mechanical Engineering, Georgia Institute of Technology, Atlanta, Georgia 30332, USA*

[2] *School of Materials Science and Engineering, Georgia Institute of Technology, Atlanta, Georgia 30332, USA*

April 9, 2017

The spectral characteristics of near-field thermal emission from nanoparticle arrays are explained by comparison to the dispersions for propagating modes. Using the coupled dipole model, we analytically calculate the spectral emission from single particles, chains, planes, and 3D arrays of $SiO_2$ and SiC. We show that the differences in their spectra are due to the existence or absence of propagating surface phonon polariton modes and that the emission is dominated by these modes when they are present. This work paves the way for understanding and control of near-field radiation in nanofluids, nanoparticle beds, and certain metamaterials.

At distances from a surface comparable to the characteristic thermal wavelength, near-field effects become significant and Planck's theory of thermal radiation is no longer valid [1,2]. When two materials are separated by such distances, evanescent surface waves such as surface phonon polaritons (SPhPs) and surface plasmon polaritons (SPPs) can tunnel between them, opening additional modes for heat transfer [3]. Many researchers have exploited this phenomenon to observe near-field radiation greater than the blackbody limit [4-12], which holds promise for applications in energy conversion, thermal management of electronics, near-field imaging, and nanomanufacturing [13].

When surface modes are spatially confined to structures such as nanopillars or nanoparticles they are considered "localized" SPhPs or SPPs. Coupling of these modes between neighboring structures could also lead to substantial energy/heat transport, which has been theoretically explored by several researchers [14-24] and recently demonstrated experimentally in packed beds of silica nanoparticles [25]. To predict the SPhP or SPP contribution to thermal transport in systems of many nanoparticles, two methodologies have emerged: calculating heat transfer by propagating SPhP or SPP waves along nanoparticle chains [14-19] and calculating particle-to-particle radiative exchanges considering many-body interactions [20-24]. It has also been suggested that the dispersion typically used for propagating SPhPs may play an important role in determining the modes of heat transfer in the absence of propagating waves [25]. Similar types of models exist for planar materials, and Shchegrov et al. [26] revealed their complementary nature by showing how propagating modes determine the spectrum of near-field thermal emission calculated from the fluctuation dissipation theorem. No such studies, however, have been performed for nanoparticle systems, and it remains unclear how collective effects may influence near-field emission from collections of nanoparticles. In this Letter, we calculate the spectrum of near-field thermal emission from ordered arrays of polaritonic nanoparticles and compare it to the density of states found from dispersion relations. Contrary to planar materials, we demonstrate that propagating modes lower and broaden the resonant peak in thermal emission, and we show that the emission characteristics are very different when propagating modes are present (in SiC systems) or absent (in $SiO_2$ systems).

We consider $N$ nonmagnetic spherical nanoparticles of radius $a$ much smaller than the characteristic thermal wavelength and center-to-center spacing $d \geq 3a$, such that they may be modeled as electric dipoles [17,27-29]. The nanoparticles are immersed in a transparent medium with relative permittivity $\varepsilon_m$. Following the example of Messina et al. [23], the spectral electric field at a location **r** emitted by the particles may be written in terms of the dyadic Green's functions as

$$\mathbf{E}(\mathbf{r}) = \mu_o \omega^2 \sum_j \mathbb{G}^{(0)}(\mathbf{r}, \mathbf{r}_j) \mathbf{p}_j(\mathbf{r}_j) \quad (1)$$

where $\mathbf{p}_j$ is the dipole moment of the $j$th nanoparticle. The free-space Green's function is

$$\mathbb{G}^{(0)}(\mathbf{r}, \mathbf{r}_j) = \frac{\exp(ik\rho)}{4\pi\rho}\left[\left(1 + \frac{ik\rho - 1}{k^2\rho^2}\right)\mathbb{1} + \frac{3(1 - ik\rho) - k^2\rho^2}{k^2\rho^2}\hat{\rho}\otimes\hat{\rho}\right] \quad (2)$$

where $k = \omega\sqrt{\varepsilon_m}/c$, $\boldsymbol{\rho} = \mathbf{r} - \mathbf{r}_j$, and $\rho = |\boldsymbol{\rho}|$. Each dipole moment $\mathbf{p}_j = \mathbf{p}_j^{(\text{fl})} + \mathbf{p}_j^{(\text{ind})}$ is split into a part due to thermal fluctuations and a part induced by the fields from other nanoparticles. The induced part can be written using the coupled dipole method [30] as

$$\mathbf{p}_j^{(\text{ind})} = \varepsilon_0 \alpha_j \mu_0 \omega^2 \sum_{k \neq j} \mathbb{G}_{jk}^{(0)} \mathbf{p}_k \quad (3)$$

Here $\alpha_j$ is the nanoparticle polarizability and the location dependence $\mathbb{G}^{(0)}(\mathbf{r}_j, \mathbf{r}_k)$ is written as $\mathbb{G}_{jk}^{(0)}$ for brevity. By combining and rewriting these equations in matrix form, we can express the total dipole moments of all particles as

$$\begin{pmatrix} \mathbf{p}_1 \\ \vdots \\ \mathbf{p}_N \end{pmatrix} = \mathbb{A}^{-1} \begin{pmatrix} \mathbf{p}_1^{(\text{fl})} \\ \vdots \\ \mathbf{p}_N^{(\text{fl})} \end{pmatrix} \quad (4)$$

The $3N \times 3N$ matrix $\mathbb{A}$ has $N \times N$ elements $\mathbb{A}_{jk}$ with $j, k = 1, \ldots, N$ and each $\mathbb{A}_{jk}$ consisting of a $3 \times 3$ tensor defined by

$$\mathbb{A}_{jk} = \delta_{jk}\mathbb{1} - (1 - \delta_{jk})\frac{\omega^2}{c^2}\alpha_j \mathbb{G}_{jk}^{(0)} \quad (5)$$

Inserting Eq. (4) into Eq. (1) yields the electric field produced by a collection of nanoparticles in terms of the fluctuating part of their dipole moments

$$\mathbf{E}_\mathbf{r} = \sum_j \mathbb{B}_{\mathbf{r}j} \mathbf{p}_j^{(\text{fl})} \quad (6)$$

where

$$(\mathbb{B}_{\mathbf{r}1} \quad \ldots \quad \mathbb{B}_{\mathbf{r}N}) = \mu_0 \omega^2 (\mathbb{G}_{\mathbf{r}1}^{(0)} \quad \ldots \quad \mathbb{G}_{\mathbf{r}N}^{(0)})\mathbb{A}^{-1} \quad (7)$$

We can now write the quantity of interest, the spectral density of emitted energy by the collection of nanoparticles, as

$$u = \frac{\varepsilon_m \varepsilon_0}{2}\langle \mathbf{E}_\mathbf{r} \cdot \mathbf{E}_\mathbf{r}^\star \rangle = \frac{\varepsilon_m \varepsilon_0}{2}\sum_{jk}\sum_{\beta\gamma}\mathbb{B}_{\mathbf{r}j,\alpha\beta}\mathbb{B}_{\mathbf{r}k,\alpha\gamma}^\star \langle p_{j,\beta}^{(\text{fl})} p_{k,\gamma}^{(\text{fl})\star}\rangle \quad (8)$$

Here the Greek letters are indices 1, 2, 3 and represent the Cartesian components of the vector quantities. The fluctuation dissipation theorem gives the relation between the fluctuating dipole moments [21]

$$\langle p_{j,\beta}^{(\text{fl})} p_{k,\gamma}^{(\text{fl})\star}\rangle = 2\frac{\varepsilon_0}{\omega}\text{Im}(\alpha_j)\Theta(\omega, T_j)\delta_{jk}\delta_{\beta\gamma} \quad (9)$$

where the mean energy of a harmonic oscillator is $\Theta(\omega, T) = \hbar\omega/[\exp(\hbar\omega/k_B T) - 1]$. The fluctuation dissipation theorem allows us to obtain a final expression for the spectral density of emitted energy

$$u(\mathbf{r}, \omega) = \frac{\varepsilon_m \varepsilon_0^2}{\omega}\sum_j [\text{Tr}(\mathbb{B}_{\mathbf{r}j}\mathbb{B}_{\mathbf{r}j}^\dagger)\text{Im}(\alpha_j)\Theta(\omega, T_j)] \quad (10)$$

This equation gives an explicit analytical method of calculating the fields at any location produced by arbitrarily arranged groups of nanoparticles knowing only their dielectric functions (expressed through the polarizability), temperatures, and the permittivity of the medium. It should be noted, however, that the size of matrix $\mathbb{A}$ scales as $(3N)^2$, which leads to computational challenges for large, three-dimensional particle systems.

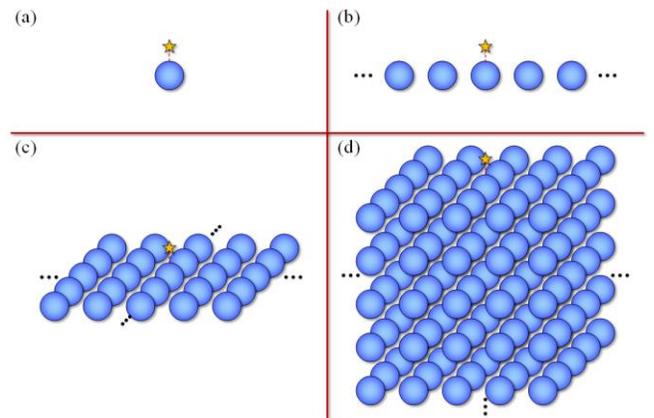

FIG. 1. Nanoparticle arrangements used for calculating near-field thermal emission. (a) Single particles, (b) chains, (c) planes, and (d) 3D crystals with spacing $d = 3a$ are considered. In each case, the emitted spectral energy density is calculated one diameter above the center particle as indicated by the star. Particles were added to each system (b-d) until the resulting spectra converged with less than 1% change at every frequency.

Nanoparticle arrays of $SiO_2$ and SiC in four different arrangements, as shown in Fig. 1, are analyzed in this work. Single particles, chains, planes, and simple cubic three-dimensional systems are considered. The radius and spacing of all nanoparticles are $a = 25$ nm and $d = 75$ nm, and all particles are at a constant temperature of 500 K. The location for the calculated local energy density is one diameter above the middle of the center-top nanoparticle as indicated by the stars in Fig. 1. To select the number of particles used for each material and geometry, particles were repeatedly added to the system until the resulting spectra changed by less than 1% for every frequency. Both $SiO_2$ and SiC support SPhPs with resonances in the Reststrahlen band between the transverse and longitudinal optical phonon frequencies. For amorphous $SiO_2$ there are two Reststrahlen bands, and the dielectric function is modeled using data from Palik [31] with $\omega_{TO,1} = 87$ Trad/s, $\omega_{LO,1} = 95$ Trad/s, $\omega_{TO,2} = 202$ Trad/s, and $\omega_{LO,2} = 234$ Trad/s. For SiC, the dielectric function is well-described by the Lorentz model

$$\varepsilon(\omega) = \varepsilon_\infty \left(1 + \frac{\omega_{LO}^2 - \omega_{TO}^2}{\omega_{TO}^2 - \omega^2 - i\omega\Gamma}\right) \quad (11)$$

with $\varepsilon_\infty = 6.7$, $\omega_{TO} = 149$ Trad/s, $\omega_{LO} = 182$ Trad/s, and $\Gamma = 0.892$ Trad/s [31]. With the dielectric function and the permittivity of the surrounding medium, each nanoparticle's polarizability can be expressed as [27]

$$\alpha_0(\omega) = 4\pi\varepsilon_m a^3 \left(\frac{\varepsilon(\omega) - \varepsilon_m}{\varepsilon(\omega) + 2\varepsilon_m}\right) \quad (12)$$

which is corrected for radiation damping by

$$\alpha(\omega) = \left(\frac{1}{\alpha_0(\omega)} - \frac{ik^3}{6\pi\varepsilon_m}\right)^{-1} \quad (13)$$

We have also checked the quadrupole contributions to polarizability [24] and found these to be negligible for all systems considered here.

With these geometries and dielectric functions in hand, the spectral energy density of emitted near-field thermal radiation is calculated with Eq. (10) and shown in Fig. 2 for particles in a medium with $\varepsilon_m = 4$, which was selected by maximizing the emission in the SiC case. For $SiO_2$ nanoparticles, resonances are seen in the two Reststrahlen bands, and the peaks rise as dimensions are added to the nanoparticle arrays. The increase and broadening of the peaks suggest additive effects and the absence of long-range coupling as more particles are added around the central location, which is further supported by low convergence requirements: the spectra converged for a 5-particle chain, $7 \times 7$-particle plane, and a $7 \times 7 \times 4$-particle 3D array. In contrast, the thermal emission from SiC nanoparticles changes significantly as the system is changed from a single particle to a chain, plane, or 3D array of particles. The peak height reduces by about 1/2 of its single-particle value and splits to form multiple peaks covering a broader spectral region. Additionally, SiC has much higher convergence requirements: 15 particles in a chain, $17 \times 17$ particles in a plane, and $17 \times 17 \times 9$ particles in a 3D array. This indicates long-range coupling and dispersive type behavior that is absent in the SiO$_2$ nanoparticle systems.

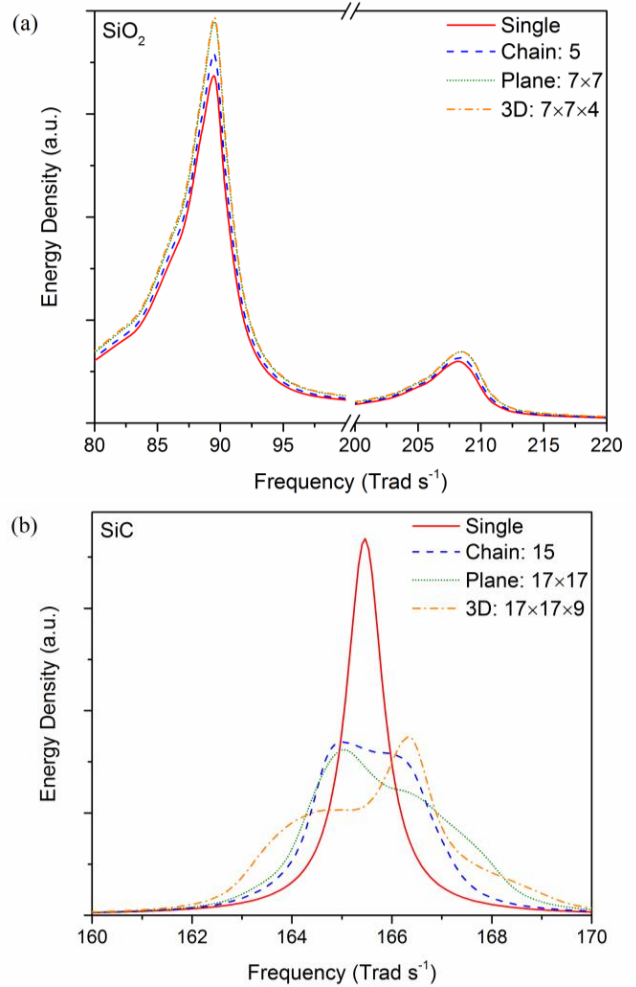

FIG. 2. Spectral energy density of emitted near-field thermal radiation from (a) $SiO_2$ nanoparticles and (b) SiC nanoparticles in a medium with $\varepsilon_m = 4$. Geometries of particle arrangements correspond to those shown in Fig. 1. The increase and broadening of peaks for $SiO_2$ suggests additive effects without strong coupling, while the change in geometry and higher convergence requirements (shown in the legends) for SiC indicates long-range coupling and dispersive behavior.

To better understand the differences between these materials and confirm that the emission characteristics of many-particle SiC systems are due to dispersive effects, we compare these spectra to the SPhP dispersion relations for nanoparticle chains. For an infinite chain of nanoparticles acting as point dipoles, the dispersion relation can be written in polylogarithm form as [32]

$$0 = 1 + \frac{\alpha}{4\pi\varepsilon_m d^3} \times$$

$$\left\{ \left[ Li_3\left(e^{i(\omega/v+\tilde{k})d}\right) + Li_3\left(e^{i(\omega/v-\tilde{k})d}\right) \right] \\ - \frac{i\omega d}{v}\left[ Li_2\left(e^{i(\omega/v+\tilde{k})d}\right) + Li_2\left(e^{i(\omega/v-\tilde{k})d}\right) \right] \\ - \left(\frac{\omega d}{v}\right)^2 \left[ Li_1\left(e^{i(\omega/v+\tilde{k})d}\right) + Li_1\left(e^{i(\omega/v-\tilde{k})d}\right) \right] \right\} \quad (14)$$

and

$$0 = 1 - \frac{\alpha}{2\pi\varepsilon_m d^3} \times$$

$$\left\{ \left[ Li_3\left(e^{i(\omega/v+\tilde{k})d}\right) + Li_3\left(e^{i(\omega/v-\tilde{k})d}\right) \right] \\ - \frac{i\omega d}{v}\left[ Li_2\left(e^{i(\omega/v+\tilde{k})d}\right) + Li_2\left(e^{i(\omega/v-\tilde{k})d}\right) \right] \right\} \quad (15)$$

for the transverse and longitudinal modes, respectively, where $d$ is the center-to-center spacing, $v = c/\sqrt{\varepsilon_m}$, $\tilde{k}$ is the complex SPhP wavevector, and $Li_n(z)$ is the polylogarithm of order $n$. Because the wavevector is taken as complex, the imaginary part yields the propagation length (1/$e$ length) $\Lambda = [2\Im(\tilde{k})]^{-1}$ of the SPhP waves traveling along the chain. These polylogarithm dispersion relations can be solved numerically with computational packages such as MATLAB, and the resulting dispersions and propagation lengths for chains of 50 nm diameter SiO$_2$ and SiC with $d = 75$ nm in $\varepsilon_m = 4$ are plotted in Fig. 3. The dispersions for SiO$_2$ in Fig. 3(a) and SiC in Fig. 3(c) do not immediately indicate a strong difference between these materials, but their propagation lengths differ significantly. For SiO$_2$ as shown in Fig. 3(b), propagation lengths are shorter than the center-to-center particle spacing, demonstrating that these modes cannot exist because they do not propagate along the chain. For SiC as shown in Fig. 3(d), however, the longest propagation lengths span multiple particles. The difference in propagation lengths is due to the higher damping (shorter lifetime) of SiO$_2$ compared to SiC [31,33]. This confirms that chains of SiC particles do indeed display dispersive SPhP behavior as indicated by the calculations of near-field thermal emission.

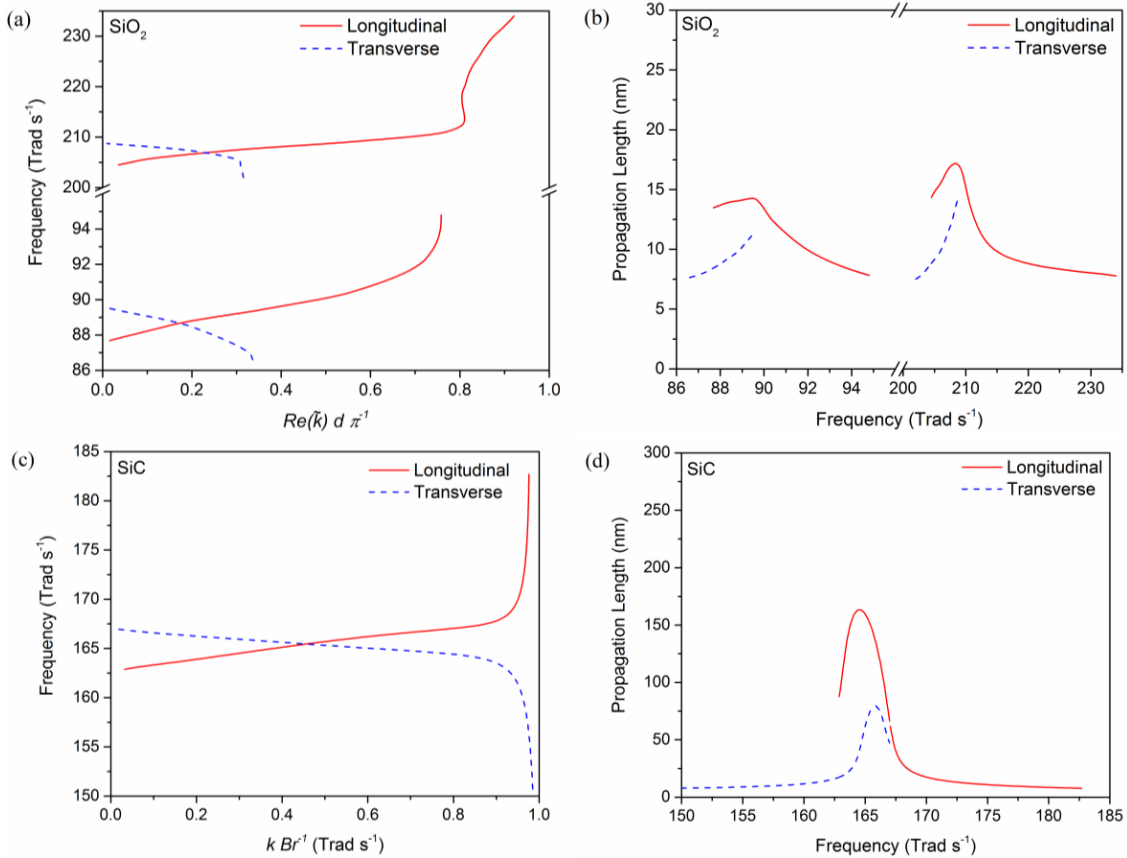

FIG. 3. Dispersion relations (a, c) and propagation lengths (b, d) for infinite chains of SiO$_2$ (a, b) and SiC (c, d) nanoparticles in $\varepsilon_m = 4$. Particles are 50 nm diameter with spacing $d = 75$ nm. The propagation lengths for SiO$_2$ are smaller than the center-to-center particle spacing, indicating that these modes cannot propagate. SiC, however, exhibits propagation lengths that span multiple particles.

To further investigate the relation between the spectral energy density of thermal emission and propagating SPhPs in SiC particle systems, we compare the density of states ($DOS$) calculated from the dispersion relation, which represents only the propagating SPhP modes, to the total emission spectrum given by Eq. (10), which represents the propagating SPhP modes and all other modes present. Because $DOS$ describes the number of modes supported at a certain frequency, one would expect a high $DOS$ where there is a peak in thermal emission if propagating SPhPs drive the emission characteristics. $DOS$ is calculated directly from the dispersion relation as $DOS = [\pi\, \partial\omega/\partial Re(\tilde{k})]^{-1}$. There are two degenerate transverse modes and one longitudinal mode, so we use $DOS = 2(DOS_{trans}) + DOS_{long}$ and plot both $DOS$ and the spectral energy density of thermal emission from Eq. (10) in Fig. 4 for SiC particle chains. Three different medium permittivities are compared to see if the thermal emission tracks changes in $DOS$. We see excellent agreement in both the peak shapes and locations, which strongly supports our initial hypothesis that the emission from SiC particle arrays is dominated by propagating SPhP modes.

It is not surprising that the characteristics of collective thermal emission in the near-field from nanoparticle arrays is largely driven by propagating SPhP modes when they are present; the same is true for emission in the near-field from flat surfaces [26]. However, propagating SPhPs on flat surfaces result in a single, sharp peak in thermal emission due to the very high $DOS$ at almost a single frequency. For ordered nanoparticle arrays on the other hand, we have shown that propagating modes lead to splitting and a decrease in intensity of the resonance peak in thermal emission compared to single nanoparticles. These spectral characteristics are due to interference and phase effects between neighboring particles governed by the dispersion of propagating SPhP waves. When propagating modes are absent, as in SiO$_2$ nanoparticle arrays, collective effects are additive around the resonance frequency of a single particle. Interference effects no longer cause a splitting of the resonance peaks, because long-range modes are not supported and the emissions from neighboring particles do not strongly influence each other. This leads to an increase in energy density at that frequency and an effective increase in $DOS$ of the system. These results lend insight to the radiative heat transfer mechanisms in nanofluids, nanoparticle beds, and metamaterials that contain repeated resonators, and they could help explain recent experimental results [25]. Much of the past research has focused on heat transfer by propagating modes only, but particle-to-particle radiation may become much more important when propagating modes are absent. Because polaritonic nanoparticles could also be used as near-monochromatic IR sources [34], these results also give an important understanding of how their emission spectra changes when collective effects become important.

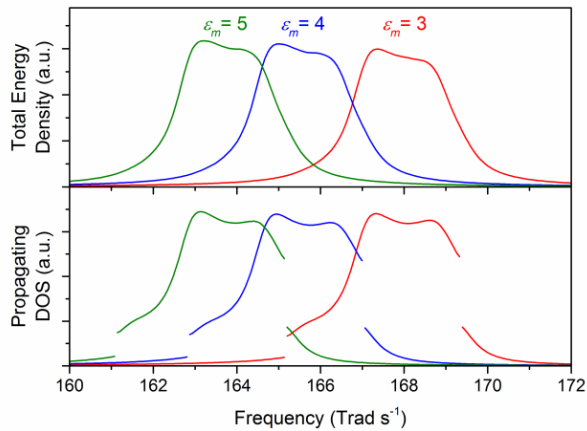

FIG. 4. Total spectral energy density of near-field thermal emission (top) compared with the density of states ($DOS$) of propagating SPhPs from the dispersion relation (bottom) for chains of SiC with $a = 25$ nm and $d = 75$ nm for three medium permittivities. $DOS$ is taken as $2(DOS_{trans}) + DOS_{long}$ because there are two degenerate transverse modes. Breaks in the $DOS$ lines are due to the fact that the transverse and longitudinal dispersions span different frequency ranges (see Fig. 3c). The excellent qualitative agreement indicates that near-field thermal emission from SiC chains is dominated by propagating SPhPs.

In summary, we have illustrated that the near-field emission spectrum from ordered nanoparticle arrays is dictated by the density of states of propagating SPhPs when they are present. Contrary to planar surfaces, propagating modes decrease and broaden the peak of thermal emission from nanoparticle arrays, and the absence of propagating modes causes an increase in the peak of thermal emission. Additional work should focus on understanding whether disorder can disrupt these propagating modes to gain better control over thermal emission as well as investigate the impacts of nanoparticle size and multipolar effects.

We would like to thank the Georgia Tech Partnership for an Advanced Computing Environment for their support with high performance computing, and we would like to thank our anonymous reviewers for their suggestions to improve this work. This material is based upon work supported by the National Science Foundation Graduate Research Fellowship Program under Grant No. (DGE-1650044). Any opinions, findings, and conclusions or recommendations expressed in this material are those of the authors and do not necessarily reflect the views of the National Science Foundation. Z.Z. also acknowledges support from the National Science Foundation (CBET-1603761).

---

*cola@gatech.edu